**Coexistence of large anomalous Nernst effect and large coercive force in amorphous ferrimagnetic TbCo alloy films**

Miho Odagiri(小田切美穂)[1], Hiroto Imaeda(今枝寛人)[1], Ahmet Yagmur[1,2], Yuichiro Kurokawa(黒川雄一郎)[3], Satoshi Sumi(鷲見聡)[1], Hiroyuki Awano(粟野博之)[1], and Kenji Tanabe(田辺賢士)[1*]
[1]Toyota Technological Institute, Nagoya, 468-8511, Japan
[2]University of Leeds, Leeds, LS2 9JT,UK
[3]Graduate School and Faculty of Information Science and Electrical Engineering, Kyushu University, Fukuoka, 819-0395, Japan
*electronic mail: tanabe@toyota-ti.ac.jp

**Abstract (199 words)**

The Anomalous Nernst Effect (ANE) has garnered significant interest for practical applications, particularly in energy harvesting and heat flux sensing. For these applications, it is crucial for the module to operate without an external magnetic field, necessitating a combination of a large ANE and a substantial coercive force. However, most materials exhibiting a large ANE typically have a relatively small coercive force. In our research, we have explored the ANE in amorphous ferrimagnetic TbCo alloy films, noting that the coercive force peaks at the magnetization compensation point (MCP). We observed that transverse Seebeck coefficients are amplified with Tb doping, reaching more than 1.0 μV/K over a wide composition range near the MCP, which is three times greater than that of pure Co. Our findings indicate that this enhancement is primarily due to direct conversion, a product of the transverse thermoelectric component and electrical resistivity. TbCo films present several significant advantages for practical use: a large ANE, the capability to exhibit both positive and negative ANE, the flexibility to be deposited on any substrate due to their amorphous nature, a low thermal conductivity, and a large coercive force. These attributes make TbCo films a promising material for advancing ANE-based technologies.

## Introduction

The Anomalous Nernst Effect (ANE) is garnering considerable interest for practical applications, notably in energy harvesting[1-3] and heat flux sensing[4]. Its ability to convert waste heat into electrical energy positions it as a potential autonomous power source for numerous sensors in Internet of Things (IoT) societies. Distinct from the Seebeck effect, the ANE generates a thermal electromotive force ($\boldsymbol{E}$) perpendicular to both the temperature gradient ($\boldsymbol{\nabla}T$) and magnetization ($\boldsymbol{M}$). The ANE is expressed as

$$\boldsymbol{E} = S_{\mathrm{ANE}}\left(\frac{\boldsymbol{M}}{|\boldsymbol{M}|} \times (-\boldsymbol{\nabla}T)\right),$$

where $S_{\mathrm{ANE}}$ is a transverse Seebeck coefficient. The conversion efficiency of ANE for energy harvesting is quantified by the dimensionless figure of merit ZT:

$$ZT = \frac{{S_{\mathrm{ANE}}}^2 \sigma_{\perp}}{\kappa_{\parallel}} T,$$

with $\sigma_{\perp}$ representing the electrical conductivity perpendicular to the temperature gradient and $\kappa_{\parallel}$ denoting the thermal conductivity parallel to the temperature gradient. Suitable materials for this purpose should, therefore, possess a large transverse Seebeck coefficient, a high electrical conductivity, and a low thermal conductivity. Furthermore, heat flux sensors, which detect heat flow, present another promising area for ANE application. These sensors are evaluated by the ratio of the thermal electromotive force $E$ to the heat flux density $j_Q$,

$$\frac{E}{j_Q} = \frac{S_{\mathrm{ANE}}}{\kappa_{\parallel}},$$

implying the necessity for a large transverse Seebeck coefficient and low thermal conductivity in these sensors. Additionally, for effective modularization, especially in IoT contexts, these materials should operate independently of a magnetic field. Thus, a large coercive force is also an essential characteristic for these applications.

The transverse Seebeck coefficient, $S_{\mathrm{ANE}}$, is a crucial parameter in ANE research. Historically, $S_{\mathrm{ANE}}$ was understood to be approximately proportional to the saturation magnetization in ordinary ferromagnetic materials. However, a pivotal moment in ANE research was the discovery by Ikhlas et al. in 2017[5] of a relatively large $S_{\mathrm{ANE}}$ value of 0.6 μV/K in antiferromagnetic $Mn_3Sn$. This discovery was particularly notable because the enhancement of $S_{\mathrm{ANE}}$ was attributed to the increased Berry curvature—an effective magnetic field in momentum space—at Weyl points near the Fermi energy. Following this study, numerous groups have identified Weyl magnetic materials with even larger $S_{\mathrm{ANE}}$ values[6-19], leading to repeated setting of new records. For instance, Sakai et al.[6] and Guin et al.[7] observed a giant ANE in the full-Heusler ferromagnet $Co_2MnGa$, with $S_{\mathrm{ANE}}$ reaching 6 μV/K at room temperature. Furthermore, Asaba et al. reported an immense ANE thermopower of $S_{\mathrm{ANE}} = 23$ μV/K in $UCo_{0.8}Ru_{0.2}Al$ at 40 K, which currently stands as the record value[9]. Despite these remarkable discoveries in recent years, a significant challenge remains: the materials identified typically exhibit extremely small coercive forces. This characteristic limits their practical application in devices where a stable magnetic response is essential.

Rare-earth (RE) and transition-metal (TM) alloys, such as TbFeCo, GdFe, and TbCo, are well-known for possessing a large coercive force[20-23]. These ferrimagnetic materials are characterized by an antiparallel alignment of the magnetic moments of TM and RE elements. Depending on the composition, there can be RE-rich films, where the net magnetization aligns with the magnetic moment of the rare-earth metal, and TM-rich films, where the transition metal predominantly contributes to the net magnetization. The boundary composition between these RE-rich and TM-rich alloys is known as the magnetization compensation point (MCP). At or near the MCP, these RE-TM alloys exhibit a significantly large coercive force, which is expected to diverge at the MCP. Additionally, due to their amorphous nature, these alloys can be deposited on various substrates at room temperature.

Ando et al. were the first to report the ANE in RE-TM alloys[24-25]. They investigated the setup arrangement's influence on ANE[25] and found transverse Seebeck coefficients of 0.08 μV/K and 0.28 μV/K in $Tb_{26}(FeCo)_{74}$ and $Tb_{36}(FeCo)_{64}$, respectively, which are RE-rich samples[24]. Seki et al. explored the anomalous Ettingshausen effect, a counterpart to the ANE, in GdCo alloys[26]. Their findings suggest that the anomalous Ettingshausen effect is not strongly influenced by the Gd composition, with the effects of Gd doping being somewhat limited. They estimated the transverse Seebeck coefficient to be 0.18 μV/K in $Gd_{22}Co_{78}$. Yagmur et al. explored the ANE in the studies on the spin Hall effect in TbCo and clarified the ANE polarity is unrelated to the polarity of the spin Hall effect[20,22]. Liu et al. studied the ANE in $Gd_{16}Co_{84}$ and $Gd_{26}Co_{74}$ alloys[27], discovering that the polarity of the ANE signal is dictated by the magnetization orientation of the Co sub-lattices rather than the overall net magnetization of GdCo. More recently, Kobayashi et al. systematically analyzed the ANE in GdFe alloys and reported a maximum transverse Seebeck coefficient value of nearly 0.3 μV/K[28]. Given that the ANE in RE-TM alloys is highly dependent on the composition ratio of RE and TM elements, as well as the proportion of Fe and Co, systematic investigations into this composition dependence are crucial. However, such comprehensive studies are currently limited.

Here, we present a comprehensive study on the composition dependence of the ANE in TbCo alloy films, examining 15 different compositions. Our findings indicate that doping Tb into the alloy significantly enhances the transverse Seebeck coefficient $S_{\mathrm{ANE}}$ compared to pure Co. Notably, $S_{\mathrm{ANE}}$ reaches more than 1.0 μV/K over a wide composition range near the MCP, with a notable change in its polarity at this point. This change suggests a direct correlation with the orientation of the Co moment. In addition to $S_{\mathrm{ANE}}$, we explored other transport parameters including the longitudinal Seebeck coefficient, electrical resistivity, and Hall resistivity. These investigations aimed to elucidate the reasons behind the observed enhancement in $S_{\mathrm{ANE}}$. Our results clarify that the enhancement is primarily due to a direct conversion mechanism $S_1$, where the temperature gradient is converted into transverse current, rather than the indirect conversion via the Seebeck effect and anomalous Hall effect $S_2$. Furthermore, TbCo films near the MCP exhibit a significant coercive force. Some samples demonstrated a combination of a large transverse Seebeck coefficient (exceeding 1.0 μV/K) and a substantial coercive force (over 3 kOe). These characteristics highlight the potential of TbCo films in applications where both large ANE and coercive force are required.

## Experiment

$Si_3N_4$(10 nm) /$Tb_xCo_{100-x}$(20 nm) /$Si_3N_4$(3 nm) films were simultaneously deposited on both $SiO_2$ glass substrates and Si substrates with thermally oxidized layers, using magnetron rf and dc sputtering techniques. For the glass samples, a Hall bar structure was formed using a metal mask, as depicted in Fig. 1(a). The TbCo alloys, varying in Tb concentration from 0 to 40, were created through a co-sputtering method. The compositions of these alloys were verified via energy dispersive X-ray spectroscopy. The transverse Seebeck coefficient, $S_{\mathrm{ANE}}$, was measured under conditions of a perpendicular magnetic field and an in-plane thermal gradient, using the samples prepared on glass substrates (as shown in Fig. 1(a)). The temperature gradient was controlled using a heater and monitored with a T-type thermocouple (copper-constantan) attached to the sample holders. While there was a minor difference between the thermocouple-measured temperature gradient and the actual gradient within the sample, this discrepancy was calibrated using thermography. To ensure the accuracy of our measurement setup, we also measured the transverse Seebeck coefficient in a Py film produced by magnetron sputtering, obtaining a value of +0.53 μV/K, consistent with previous findings[29]. The polarity of $S_{\mathrm{ANE}}$, as defined in Fig. 1(a), aligns with established conventions: $S_{\mathrm{ANE}}$ is positive (>0) in materials such as Co, Ni, and Py, and negative (<0) in Fe. Additionally, the longitudinal Seebeck coefficient, electrical resistivity, and Hall resistivity were measured using the same sample, with the setups illustrated in Figs. 1(b-d). The four-terminal method was employed for electrical resistivity measurements. Kerr angle and magnetization data were acquired using a Kerr loop tracer and a vibrating sample magnetometer, respectively, with the Kerr loop tracer utilizing a 690 nm wavelength laser. For Kerr angle and magnetization measurements, Si-substrate samples featuring a 350 nm thermally oxidized layer were used. This layer enhances the Kerr angle due to multiple interference effects, in contrast to the films on the glass substrate[30].

## Results

Figures 2(a-b) present the saturation magnetization and Kerr angle as functions of the TbCo composition. The saturation magnetization decreases as the Tb concentration increases, reaching a minimum at $x = 20$. Simultaneously, the Kerr angle undergoes a polarity change at this composition, indicating that $x = 20$ is near the MCP. The Kerr angle polarity in TbCo alloys is influenced by the direction of the Co moment, suggesting that the transition metal (TM)-rich region corresponds to $x < 20$, while the rare-earth (RE)-rich region is for $x > 20$. The coercive force exhibits a marked increase around the MCP, reaching up to 6 kOe for $x = 19.9$. In Figure 2(c), the detected voltage is plotted as a function of an external magnetic field for $x = 22.7$. The appearance of a clear hysteresis curve in this plot indicates that the sample has a perpendicular magnetic anisotropy. The coercive field for this sample is approximately 3600 Oe, which is considerably large. Samples with $x$ ranging from 10 to 30 display perpendicular magnetization, while others exhibit in-plane magnetization. The ANE voltage $V$ is defined as half of the voltage difference at high magnetic fields between positive and negative saturations. This voltage $V$ is observed to increase with the rising temperature difference. Figure 2(d) illustrates the ANE electric field $E$ as a function of the temperature gradient $\nabla T$. Here, $E$ and $\nabla T$ are derived from dividing $V$ by the length 7 mm and dividing the temperature difference by the length 6 mm, respectively. $E$ shows a direct proportionality to $\nabla T$. The blue dotted line represents a linear fitting, and its slope corresponds to the transverse Seebeck coefficient $S_{\mathrm{ANE}}$.

Figure 2(e) illustrates the composition-dependent behavior of $S_{\mathrm{ANE}}$ in TbCo films. A notable change in the sign of $S_{\mathrm{ANE}}$ is observed between compositions $x = 19.9$ and 22.7, corresponding to the change from a TM-rich to an RE-rich alloy composition. It means that ANE in TbCo is predominantly influenced by the Co moment, aligning with previous findings in GdCo[26-27] and TbFeCo[24-25]. Unlike other materials with large $S_{\mathrm{ANE}}$ values, such as $Co_2MnGa$[31], $Fe_3Ga$[11], $Fe_3Al$[11], and SmCo[32], which exhibit positive polarity, corresponding to the opposite polarity in Fe and the same polarity in Py and Co, RE-rich TbCo alloys demonstrate significant negative polarity. This characteristic is crucial for applications requiring both positive and negative $S_{\mathrm{ANE}}$ values. Additionally, the magnitude of $S_{\mathrm{ANE}}$ over a wide composition range near the MCP remains relatively stable, around 1.0 μV/K, irrespective of the composition in Fig. 2(f). The presence of a large $S_{\mathrm{ANE}}$ around the MCP, coupled with the sign change at the MCP, is vital for the development of thermoelectric power generation modules. Such applications require materials capable of exhibiting both positive and negative $S_{\mathrm{ANE}}$ and a strong coercive force to function without an external magnetic field. TbCo films near the MCP meet these requirements.

In the specific case of $x$ = 19.9, close to the MCP, the hysteresis behavior deviates from the typical rectangular shape, as shown in Fig. 2(g). A peak around $\pm 6$ kOe, reminiscent of the topological Hall effect[33-36], is observed. This peak is often seen in systems where two phases with distinct coercive forces coexist[37]. A similar hysteresis pattern near the MCP temperature was reported in studies on an anomalous Hall effect in GdCo[38], attributed to the coexistence of TM-rich and RE-rich regions. Since MCP is also temperature-dependent, our TbCo sample with $x$ = 19.9 under a temperature gradient likely has both TM-rich and RE-rich regions. Figure 2(h) displays the electric field $E$ as a function of the temperature gradient $\nabla T$ for $x$ = 19.9. We analyzed the voltages $V$ at high magnetic fields (orange circle) and at the peak (pink triangle) from Fig. 2(e). Both $E$ values show a direct proportionality to $\nabla T$, indicating a consistent behavior between the TM and RE regions for each temperature difference. The pink triangles in Figs. 2(e-f) indicate the data calculated by the green slope in Fig. 2(h). Its value is smaller than the other $|S_{\mathrm{ANE}}|$ around the MCP, suggesting the existence of area with a coercive force higher than the external magnetic field.

## Discussion

We examined the composition dependencies of electrical resistivity $\rho_{xx}$, Hall resistivity $\rho_{yx}$, and longitudinal Seebeck coefficient $S_{xx}$ in TbCo alloys to understand the underlying reasons for the observed large $S_{\mathrm{ANE}}$ values. As shown in Fig. 3(a-c), $S_{xx}$ decreases with increased Tb doping and becomes notably small around the MCP. According to transport property theories, the transverse Seebeck coefficient $S_{\mathrm{ANE}}$ can be decomposed into two distinct contributions,

$$S_{\mathrm{ANE}} = \alpha_{yx}\rho_{xx} + \alpha_{xx}\rho_{yx}.$$

Here, $\rho_{xx}$ is the electrical resistivity, $\rho_{yx}$ is the Hall resistivity, $\alpha_{xx}$ is the longitudinal thermoelectric constant, and $\alpha_{yx}$ is the transverse thermoelectric constant.

The terms $\alpha_{yx}\rho_{xx}$ and $\alpha_{xx}\rho_{yx}$ correspond to $S_1$ and $S_2$, respectively. Since $\alpha_{xx}$ is defined as $S_{yx}/\rho_{xx}$, all parameters in the equation, except for $\alpha_{yx}$, can be directly measured in experiments. $S_2$ is derived from the interplay of the Seebeck effect and the anomalous Hall effect, representing an indirect conversion where electrons driven by the temperature gradient are redirected perpendicularly by the anomalous Hall effect. Conversely, $S_1$, based on $\alpha_{yx}$, represents a direct conversion mechanism where the transverse electric current is generated directly from the temperature gradient, independent of the Seebeck effect. By analyzing the $\rho_{xx}$, $S_{xx}$, and $\rho_{yx}$ data, we calculated the contribution of $S_2$. Subsequently, the contribution of $S_1$ was determined using the equation $S_1 = S_{\mathrm{ANE}} - S_2$, as illustrated in Fig. 3(d). Our findings indicate that $S_1$ is significantly larger than $S_2$, highlighting the importance of direct conversion in enhancing $S_{\mathrm{ANE}}$.

Further, we delved into the transport properties related to the anomalous Hall effect. Based on the theoretical framework proposed by Onoda et al.[39], the anomalous Hall effect can be categorized into three regimes: clean, intrinsic, and dirty. As depicted in Fig. 4(a), our analysis shows that low-doped TbCo alloys are in the intrinsic regime, while high-doped TbCo alloys align with the dirty regime. The Hall conductivity in TbCo is highly dependent on electrical conductivity, suggesting that the anomalous Hall effect in these alloys is primarily influenced by impurity scattering mechanisms, such as side jump and skew scattering. This implies that the doped Tb element acts as a scattering impurity for electrons.

Figure 4(b) encapsulates the relationship between the transverse Seebeck coefficient and saturation magnetization in TbCo alloys. Unlike ferromagnetic materials where $S_{\mathrm{ANE}}$ tends to be proportional to the saturation magnetization, Weyl magnetic materials exhibit an upward trend with saturation magnetization, although with significant deviations. Notably, TbCo shows a unique insensitivity to changes in magnetization, allowing for tailored saturation magnetization levels depending on specific applications. Figure 4(c) illustrates the correlation between $S_{\mathrm{ANE}}$ and coercive force at room temperature. Most materials with a large $S_{\mathrm{ANE}}$ exhibit a negligible coercive force, and it is rare to find materials combining both a large coercive force (over 1000 Oe) and a significant $S_{\mathrm{ANE}}$ (over 1.0 μV/K). TbCo stands out as an exceptional material in this regard, possessing both a large coercive force and a substantial $S_{\mathrm{ANE}}$.

Finally, let us summarize the characteristics of TbCo. Firstly, TbCo exhibits a large $S_{\mathrm{ANE}}$, particularly around the MCP. The alloy demonstrates both a significant coercive force and the presence of both positive and negative $S_{\mathrm{ANE}}$ near the MCP.

Notably, materials with a large negative $S_{\mathrm{ANE}}$ are scarce, making TbCo especially valuable. Secondly, TbCo exhibits amorphous and a low thermal conductivity. An amorphous material has a low thermal conductivity. According to the report by Hopkins et al.[40], the thermal conductivities in $Gd_{21}Fe_{72}Co_7$ and $Tb_{21}Fe_{73}Co_6$ are approximately 5 W/mK, which is considerably lower compared to 24 W/mK in $Co_2MnGa$[6]. The TbCo alloys also expected to have a similar thermal conductivity. This characteristic enhances its suitability for various applications. Thirdly, TbCo has versatility in substrate compatibility. TbCo can be deposited on any substrate at room temperature, offering significant flexibility for practical applications. Fourthly, the magnitude of the longitudinal Seebeck coefficient ($|S_{xx}|$) remains relatively small (< 5 μV/K) and consistent around the MCP. For heat flux sensor applications, where two materials with similar $S_{xx}$ and larger $S_{\mathrm{ANE}}$ are required, TbCo's properties are advantageous. While Tanaka et al. achieved this condition using FeGa and FeGa/CuNi[41], TbCo alloys simplify the process by utilizing TM-rich and RE-rich compositions near the MCP.

## Conclusion

Our comprehensive study on the ANE in ferrimagnetic TbCo alloy films has yielded significant insights. We found that doping Tb into TbCo alloys substantially enhances the transverse Seebeck coefficient ($S_{\mathrm{ANE}}$), achieving a value of more than 1.0 μV/K over a wide composition range near the MCP. This value is notably three times greater than that of pure Co. Additionally, the coercive force in these alloys shows a tendency to increase sharply, reaching a peak of 6 kOe near the MCP. To understand the reasons behind this enhancement, we delved into the composition dependencies of various properties, including the longitudinal Seebeck coefficient, electrical resistivity, and Hall resistivity. Our findings indicate that the increase in $S_{\mathrm{ANE}}$ is primarily due to direct conversion mechanisms. Overall, our research reveals that TbCo alloys possess several key advantages for practical applications. The substantial increase in $S_{\mathrm{ANE}}$, coupled with a significant coercive force near the MCP, makes these materials particularly suitable for a range of applications where large ANE and coercive force are crucial. This study, therefore, highlights the potential of TbCo alloys in advancing technologies that leverage the ANE.

## ACKNOWLEDGEMENT

This work was partly supported by a Grant-in-Aid for Scientific Research (C) (No. 20K05307) from JSPS and by the Paloma Environmental technology Development Foundation.

## AUTHOR DECLARATIONS

### Conflict of Interest

The authors have no conflicts to disclose.

### Author Contributions

**Miho Odigiri**: Data Curation (lead); Investigation (lead); Formal Analysis (lead); Writing/Original Draft Preparation (supporting); **Hiroto Imaeda**: Data Curation (supporting); Formal Analysis (supporting); Methodology (supporting); **Ahmet Yagmur**: Data Curation (supporting); Methodology (supporting); Writing/Review & Editing (supporting); **Yuichiro Kurokawa**: Data Curation (supporting); Writing/Review & Editing (supporting); **Satoshi Sumi**: Data Curation (supporting); Writing/Review & Editing (supporting); **Hiroyuki Awano**: Investigation (supporting); Resources (equal); Writing/Review & Editing (supporting); **Kenji Tanabe**: Conceptualization (lead); Data Curation (supporting); Formal Analysis (supporting); Funding Acquisition (lead); Investigation (supporting); Methodology (lead); Project Administration (lead); Resources (equal); Supervision (lead); Writing/Original Draft Preparation (lead); Writing/Review & Editing (lead)

## DATA AVAILABILITY

The data that support the findings of this study are available from the corresponding author upon reasonable request.

## Figure Caption

Fig. 1(a-d) Schematic illustrations of setups for measuring the ANE (a), Seebeck effect (b), Hall resistance (c) and resistance (d). All the parameters were measured by using the same sample. The four terminal method was used in the resistance measurement.

Fig. 2(a-b) Saturation magnetization(a) and Kerr angle(b) as a function of TbCo composition. $4\pi M_s$ of 18 kG as $x$ = 0.0 corresponds to pure Co. $4\pi M_s$ decreases with increasing $x$ because the Tb moment is antiparallel to the Co moment. $M_s$ is minimized and the sign of the Kerr angle is changed when $x$ = 19.9, which is close to a magnetization compensation point. The blue curves are guides to the eye. (c) ANE voltage as functions of a magnetic field and temperature difference applied into the sample as $x = 22.7$. (d) Relationship between the voltage and temperature difference as $x = 22.7$. The blue line indicates a linear fitting line and its slop corresponds to $S_{\mathrm{ANE}}$. (e-f) Transverse Seebeck coefficient(e) and its absolute value(f) as a function of TbCo composition. The pink triangles indicate the result in the sample with the composition close to a magnetization compensation point. The blue curves are guides to the eye. (g) ANE voltage as functions of a magnetic field and temperature difference as $x = 19.9$. A peak appears at 6 kOe(the green dotted circle). (h) Relationship between the voltage and temperature difference as $x = 19.9$. The pink triangle and orange circle indicate the voltages saturated under the high magnetic field and of the peak, respectively. The blue and green dotted lines indicate linear fitting lines. The pink triangles in Figs. 2(e-f) indicate the results calculated from the voltages of the peak in the sample. $S_{\mathrm{ANE}}$ is suppressed at the MCP. The origin of the suppression may be that there is an area having a coercive force larger than the maximum magnetic field.

Fig. 3(a-c) Electrical resistivity(a), Hall resistivity(b), and longitudinal Seebeck coefficient(c) as a function of TbCo composition. (d) $S_{\mathrm{ANE}}$, $S_1$, and $S_2$ as a function of TbCo composition. $S_2$ is derived by $S_{xx}\rho_{yx}/\rho_{xx}$ from Figs. 3(a-c). $S_1$ is derived by $S_{\mathrm{ANE}} - S_2$.

Fig. 4(a) Summary of experimental results on the relationship between $|\sigma_{xy}|$ and $\sigma_{xx}$ in various ferromagnets. $\sigma_{xx} = \rho_{xx}/({\rho_{xx}}^2 + {\rho_{yx}}^2)$ and $\sigma_{xy} = \rho_{yx}/({\rho_{xx}}^2 + {\rho_{yx}}^2)$. The data are obtained in Ref. 42 (Fe, Co, Ni, Gd), Ref. 43 (GaMnAs), Ref. 44 (Co-$TiO_2$), Ref. 45 (MnSi), and Ref. 46 ($Nd_2(MoNb)_2O_7$). (b) Summary of experimental results on the relationship between $|S_{\mathrm{ANE}}|$ and $4\pi M_s$ in various ferromagnets. The data are obtained in Ref. 47 (Fe, Co), Ref. 48 (FePt), Ref. 49 (MnGe), Ref. 14 (MnSn,

$Mn_{3.04}Ge_{0.96}$), Ref. 50 ($D0_{22}$-$Mn_2Ga$, $L1_0$-MnGa, $L1_0$-FePt, $L1_0$-FePd), Ref. 11 ($Fe_3Ga$, $Fe_3Al$), Ref. 6 ($Co_2MnGa$), Ref. 13 ($Co_3Sn_2S_2$), Ref. 9 (UCoRuAl), and Ref. 51 ($Fe_3O_4$). (c) Summary of experimental results on the relationship between $|S_{\mathrm{ANE}}|$ and $H_{\mathrm{c}}$ in various ferromagnets. The data are obtained in Ref. 27 (CoGd), Ref. 24 (TbFeCo), Ref. 52 ($Fe_4N$), and Ref. 53 ($Mn_4N$). When the coercive force is below the detection limit, it is assumed to be 1 Oe.

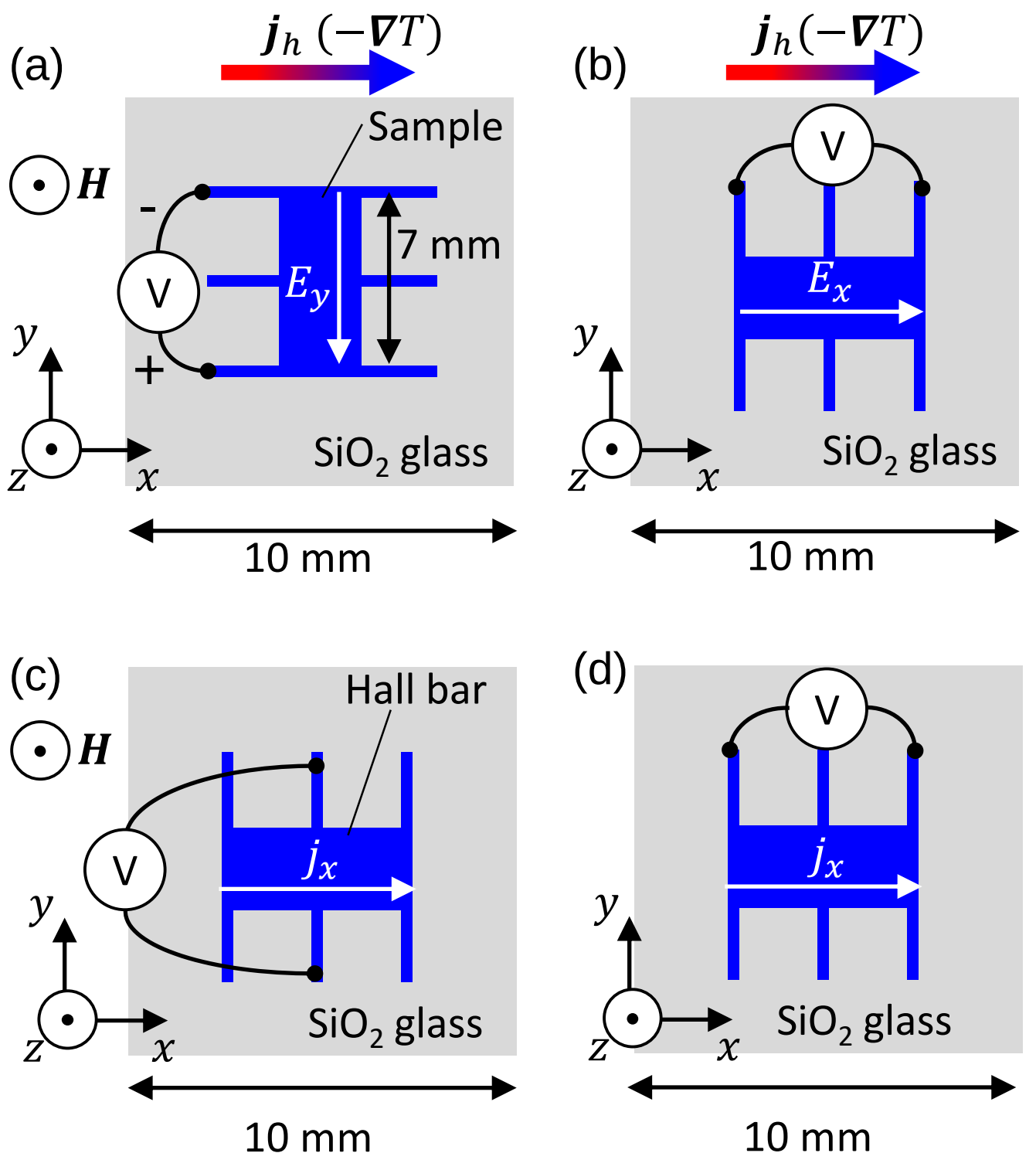

Fig. 1

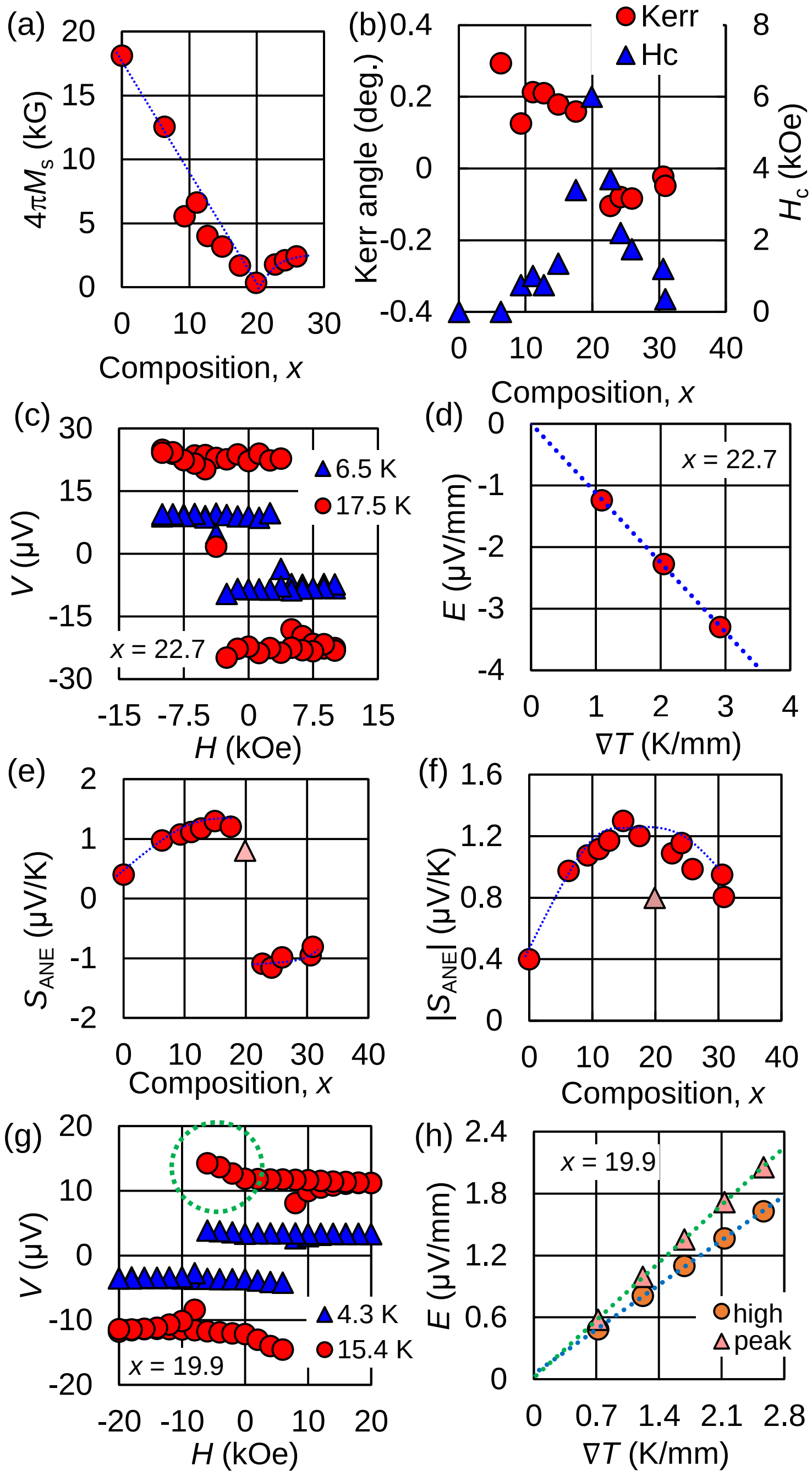

Fig. 2

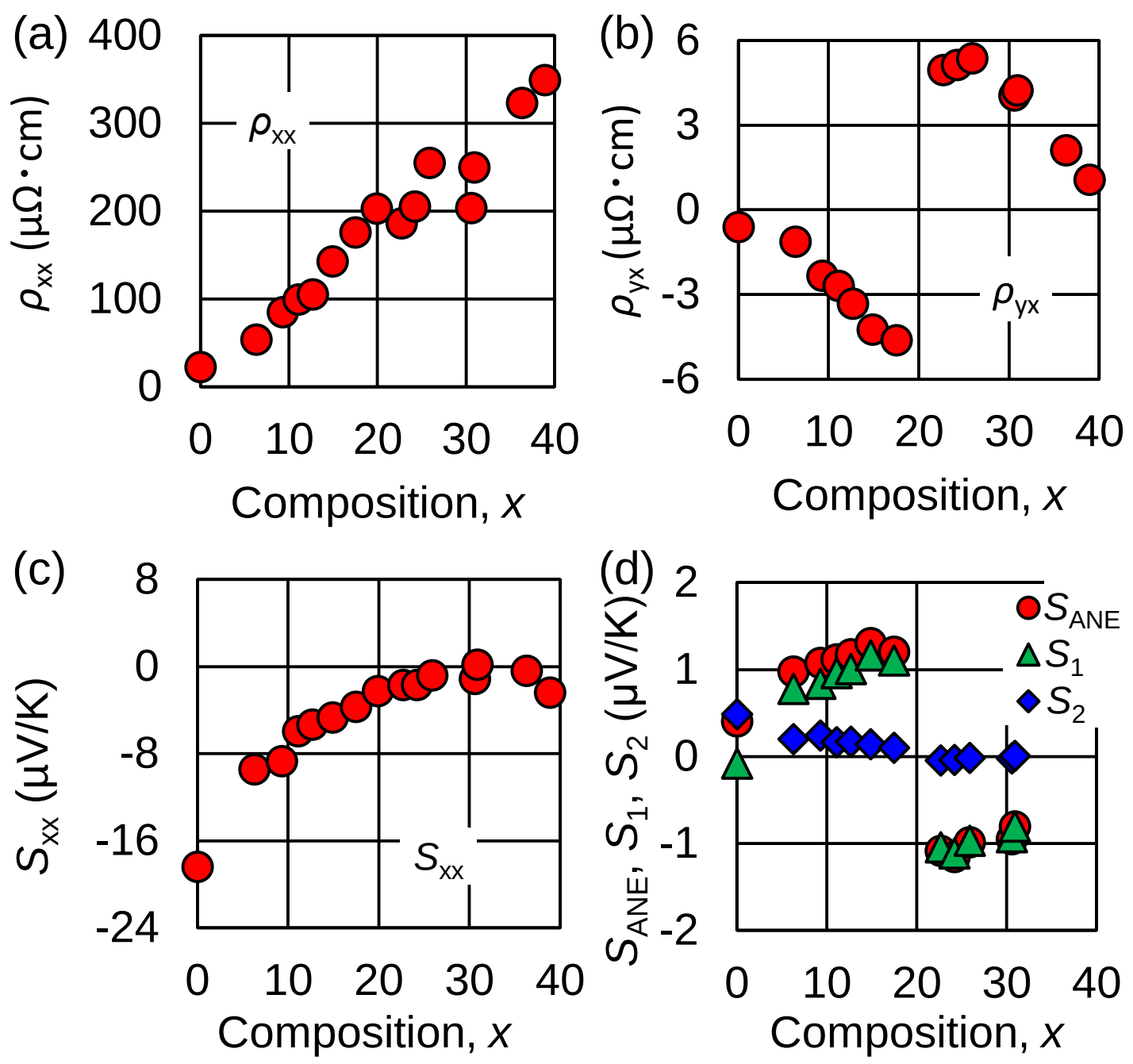

Fig. 3

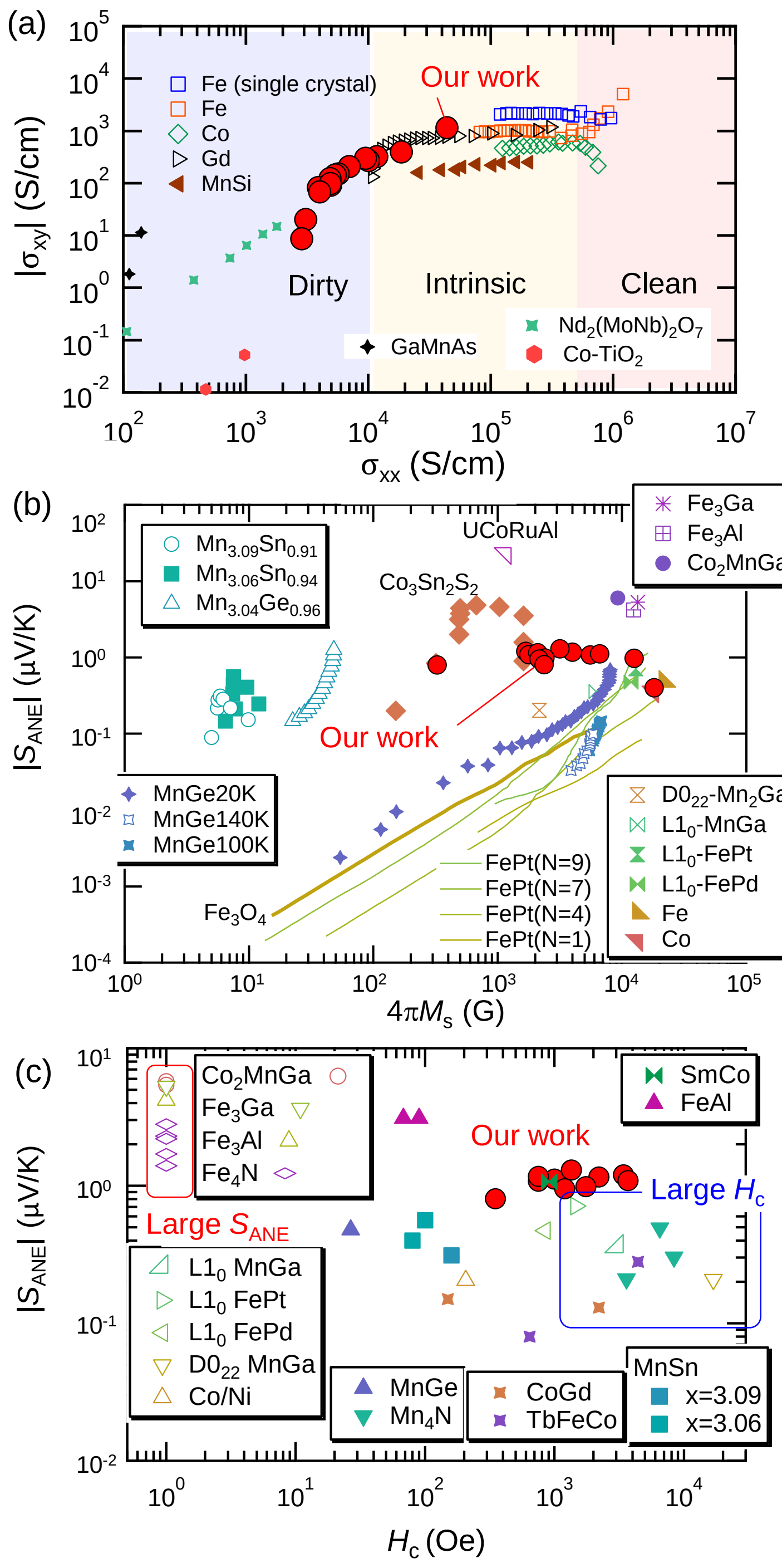

Fig. 4